\documentclass[12pt]{iopart}
\usepackage{graphicx}
\begin{document}
\newcommand{\bk}{{\bf k}}
\newcommand{\bK}{{\bf K}}
\newcommand{\tdia}{t^{\dagger}_{i\alpha}}
\newcommand{\tia}{t_{i\alpha}}
\newcommand{\tdja}{t^{\dagger}_{j\alpha}}
\newcommand{\tja}{t_{j\alpha}}
\newcommand{\tdjg}{t^{\dagger}_{j\gamma}}
\newcommand{\tjg}{t_{j\gamma}}
\newcommand{\tdig}{t^{\dagger}_{i\gamma}}
\newcommand{\tig}{t_{i\gamma}}
\newcommand{\tdib}{t^{\dagger}_{i\beta}}
\newcommand{\tib}{t_{i\beta}}
\newcommand{\tdjb}{t^{\dagger}_{j\beta}}
\newcommand{\tjb}{t_{j\beta}}
\newcommand{\tdnoa}{t^{\dagger}_{n+1,\alpha}}
\newcommand{\tnoa}{t_{n+1,\alpha}}
\newcommand{\tdng}{t^{\dagger}_{n\gamma}}
\newcommand{\tng}{t_{n\gamma}}
\newcommand{\tdnod}{t^{\dagger}_{n+1,\delta}}
\newcommand{\tnod}{t_{n+1,\delta}}
\newcommand{\tdag}{t^{\dagger}}
\newcommand{\al}{\alpha}
\newcommand{\be}{\beta}
\newcommand{\ca}{\gamma}
\newcommand{\de}{\delta}
\newcommand{\taud}{\tau^{\dagger}}
\newcommand{\bea}{\begin{eqnarray}}
\newcommand{\eea}{\end{eqnarray}}

\title{ 
Critical Behaviour of Structure Factors at a Quantum Phase Transition
}
\author{ C. J. Hamer}
\address{School of Physics, The University of New South Wales,
  Sydney, NSW 2052, Australia}
\date{\today}
\begin{abstract}
We review the theoretical behaviour of the total and one-particle structure factors at a quantum
phase transition for temperature $T=0$. The predictions are compared with exact or numerical
results for the transverse Ising model, the alternating Heisenberg chain, and the bilayer
Heisenberg model. At the critical wavevector, the results are generally in accord with
theoretical expectations. Away from the critical wavevector, however, different models display
quite different behaviours for the one-particle residues and structure factors.
\end{abstract}
\submitted{\JPCM}

\section{Introduction}
\label{sec1}

Modern probes of material properties, such as the new inelastic neutron
scattering facilities, are reaching such unprecedented sensitivity that
they can measure the spectrum not only of a single quasiparticle
excitation, but even two-particle excitations (e.g. \cite{tennant2003}).
These quasiparticles
can collide, scatter, or form bound states just like elementary
particles in free space. The spectrum of the multiparticle excitations
is a crucial indicator of the underlying dynamics of the system.

The experiments measure scattering cross-sections, which are
proportional to the appropriate 'structure factor' for the system or
material at hand \cite{marshall1971,nielsen1976}. It is therefore of particular
interest to explore the critical behaviour of these structure factors in
the vicinity of a quantum phase transition. In this paper, we present a
review of this topic, comparing the theoretical predictions with some
exact analytic results and numerical calculations for various models.
We concentrate here on quantum spin models, but the major conclusions
are applicable more generally.

The theoretical behaviour of the total structure factor has been discussed
since early days. More
recently, people have begun to discuss the breakdown of the total
structure factor into its component multiparticle contributions from one, two, .. etc.
intermediate quasiparticles.
Sachdev \cite{sachdev1999}, for instance, discusses the behaviour of the
1-particle structure factor in his book on quantum phase transitions. In
Section \ref{sec2} of the paper, we draw together these theoretical
discussions.

In the remainder of the paper, we review the behaviour of the structure
factors for some specific models. In Section \ref{sec3} we look at the
transverse Ising chain, which is exactly solvable, and hence yields some
exact results for the 1-particle structure factors \cite{hamer2006}. In Section
\ref{sec4}, we review some numerical results obtained by series
expansion methods for some other models, namely the trasverse Ising
model in higher dimensions \cite{hamer2006}. the alternating Heisenberg chain
\cite{schmidt2003,hamer2003}, and the bilayer Heisenberg antiferromagnet \cite{collins2008}.

Our main conclusions, in Section \ref{sec5}, concern the relationship
between the 1-particle structure factor and the total structure factor.
It is usually assumed that the 1-particle term dominates the total
structure factor, and their scaling behaviour is the same; but this is
not always strictly true. In the transverse Ising model and the
dimerized alternating chain, for example, te 1-particle structure factor
actually vanishes at the critical coupling, everywhere except at the
critical wavevector. Only for the bilayer model does the 1-particle
structure factor remain dominant at all wavevectors. This latter
behaviour, however, is presumably more typical in generic quantum spin
systems.
\section{Review of Theory}
\label{sec2}

Assuming magnetic scattering from atomic spins ${\bf S}_i$ localized on sites $i$ of a Bravais crystal lattice, the
neutron scattering cross section can be directly related to the dynamical structure factor \cite{marshall1971} 

\begin{equation}
S^{\alpha\gamma}({\bf k},\omega) = \frac{1}{2\pi N} \sum_{i,j} \int^{\infty}_{-\infty} dt 
\ e^{i(\omega t - {\bf k \cdot (r_j - r_i)})} C^{\alpha\gamma}({\bf r_j-r_i},t)
\label{eqB1}
\end{equation}
where
\begin{equation}
C^{\alpha\gamma}({\bf r_j-r_i},t) = <S^{\alpha}_j(t) S^{\gamma}_i(0)> .
\label{eqB2}
\end{equation}
Here $i,j$ label sites of the lattice, $\alpha,\gamma$ label Cartesian components of the spin operator ${\bf
S}$, $N$ is the number of lattice sites, $C^{\alpha\gamma}({\bf r},t)$ is the spin-spin correlation function, 
and the angular bracket denotes the thermal expectation value at
finite $T$ or, at $T=0$, the ground-state expectation value.
The structure factor satisfies the condition of `detailed balance'
\begin{equation}
S^{\alpha\gamma}({\bf k},\omega) = e^{\beta \omega} S^{\gamma\alpha}(-{\bf k},-\omega) .
\label{eqB2a}
\end{equation}
where $\beta=1/k_BT$ in the exponent is the usual Boltzmann coefficient.
The time dependence of the spin operator is given as usual by
\begin{equation}
S^{\alpha}_j(t) = e^{iHt}S^{\alpha}_j(0)e^{-iHt}
\label{eqB3}
\end{equation}

Integrating over energy gives the `integrated' or `static' structure factor
\begin{equation}
S^{\alpha\gamma}({\bf k})  =  \int^{\infty}_{-\infty} d\omega \ S^{\alpha\gamma}({\bf k},\omega) 
  =  \frac{1}{N} \sum_{i,j} e^{i{\bf k \cdot (r_i-r_j)}}<S^{\alpha}_jS^{\gamma}_i>,
\label{eqB4}
\end{equation}
the spatial Fourier transform of the 2-spin correlator at equal times.

Integrating over momentum then yields a sum rule:
\begin{equation}
\frac{1}{N} \sum_{\bf k} \int^{\infty}_{-\infty} d\omega \ S^{\alpha\gamma}({\bf k},\omega) = \frac{1}{N}\sum_i
<S^{\alpha}_i S^{\gamma}_i>,
\label{eqB5}
\end{equation}
involving the expectation value of two spin operators at the same point.

If $S^{\alpha}$ and $S^{\gamma}$ are Hermitian conjugates, which is usually the case of most interest, we can
introduce a complete set of energy eigenstates $|n>$ in equation (\ref{eqB1}) and integrate over time to get
\begin{equation}
S^{\alpha\gamma}({\bf k},\omega) = \sum_n S^{\alpha\gamma}_n({\bf k},\omega),
\label{eqB6}
\end{equation}
i.e. a sum over `exclusive' structure factors or `spectral weights' $S^{\alpha\gamma}_n$, where
\begin{equation}
S^{\alpha\gamma}_n({\bf k},\omega)  =  
\frac{1}{N}\sum_n\delta(\omega-E_n+E_0)|\sum_i<\psi_n|S^{\gamma}_i|\psi_0>e^{i{\bf k \cdot r_i}}|^2 \hspace{5mm} (T=0)
\label{eqB7}
\end{equation}
or for $T \neq 0$
\begin{equation}
S^{\alpha\gamma}_n({\bf k},\omega)   =  
\frac{1}{NZ}\sum_{m,n}\delta(\omega-E_n+E_m)e^{-\beta E_m}|\sum_i <\psi_n|S^{\gamma}_i|\psi_m>e^{i{\bf k \cdot
r_i}}|^2  
\label{eqB7a}
\end{equation}
where $E_n$ is the energy of the nth eigenstate, $|\psi_0>$ is the ground state, and $Z$ is the
partition function
\begin{equation}
Z = \sum_n e^{-\beta E_n}.
\label{eqB7a}
\end{equation}
If the system exhibits well-defined quasiparticle excitations, the intermediate states $n$ can be classified
into 1-particle, 2-particle or many-particle states, each state making a non-negative contribution, so that the
total structure factor is real and positive semi-definite.

Following Sachdev \cite{sachdev1999}, we may also define the corresponding generalized susceptibility
$\chi^{\alpha\gamma}({\bf k},\omega)$ by a Fourier transform in imaginary time ($it \rightarrow \tau$)
\begin{equation}
\chi^{\alpha\gamma}({\bf k},\omega_n) = \int^{\beta}_0 d\tau \sum_i C^{\alpha\gamma}({\bf r_i},\tau)e^{-i({\bf k
\cdot r_i -\omega_n\tau)}}
\label{eqB8}
\end{equation}
where $\omega_n = 2\pi nT$, $n$ integer, is the Matsubara frequency arising from periodic boundary conditions
across the strip of width $\beta$ in imaginary time. Then $\chi^{\alpha\gamma}({\bf k},\omega)$ for real
frequencies is obtained by an analytic continuation $i\omega_n \rightarrow \omega +i\delta$, where $\delta$ is
a positive infinitesimal. The dynamic susceptibility measures the response of the magnetization $S^{\alpha}$ to
an external field coupled linearly to $S^{\gamma}$, oscillating with wavevector ${\bf k}$ and frequency
$\omega$. 
One can show \cite{marshall1971} that
$\chi^{\alpha\gamma}$ satisfies the Kramers-Kronig relation
\begin{equation}
Re\{\chi^{\alpha\gamma}({\bf k},\omega)\} = P\int^{\infty}_{-\infty} \frac{d\Omega}{\pi} \frac{Im\{\chi^{\alpha\gamma}({\bf
k},\Omega)\}}{\Omega-\omega}
\label{eqB9}
\end{equation}
where $P$ indicates the principal part. 

If $S^{\alpha}$ and $S^{\gamma}$ are Hermitian conjugates, then a
fluctuation-dissipation theorem connects the structure factor $S^{\alpha\gamma}$ to the imaginary part of the dynamic
susceptibility \cite{marshall1971,sachdev1999}:
\begin{equation}
S^{\alpha\gamma}({\bf k},\omega) = \frac{1}{\pi(1-e^{-\beta\omega})} Im\{\chi^{\alpha\gamma}({\bf k},\omega)\}
\label{eqB10}
\end{equation}

If $S^{\alpha}$ and $S^{\gamma}$ are themselves Hermitian, one can show, using spectral analysis as for $S^{\alpha\gamma}$ above, that
\begin{equation}
\chi^{\alpha\gamma *}({\bf k},\omega) = \chi^{\alpha\gamma}(-{\bf k},-\omega)
\label{eqB10a}
\end{equation}

If both conditions are true, i.e. $\alpha = \gamma$ and $S^{\alpha}$ is
Hermitian, then the diagonal susceptibility obeys
\begin{equation}
\chi^{\alpha\alpha}({\bf k},\omega) = \chi^{\alpha\alpha}(-{\bf k},\omega)
\label{eqB10b}
\end{equation}
and
\begin{equation}
\chi^{\alpha\alpha}({\bf k},-\omega) = \chi^{\alpha\alpha *}({\bf k},\omega)
\label{eqB10c}
\end{equation}
Thus $Im\{\chi^{\alpha\alpha}\}$ is an
odd function of $\omega$, while $Re\{\chi^{\alpha\alpha}\}$ is an even function of $\omega$. 
From (\ref{eqB10}), the
dynamic structure factor then satisfies
\begin{equation}
S^{\alpha\alpha}({\bf k},-\omega) = e^{-\beta\omega}S^{\alpha\alpha}({\bf k},\omega)
\label{eqB11}
\end{equation}

\subsection{Critical Behaviour near a Quantum Phase Transition}

Now let us suppose that a quantum spin model undergoes a quantum phase transition as a function of some coupling
$\lambda$ at temperature $T=0$.
The critical behaviour of the integrated structure factor can be
obtained from a heuristic argument as follows.
In the continuum
approximation near the critical point, equation (\ref{eqB5}) for the static structure factor reduces to
\begin{equation}
S^{\alpha\gamma}({\bf k}) = \int d^d r \ e^{i{\bf k \cdot r}} <S^{\alpha}({\bf r}) S^{\gamma}(0)>_0
\label{eqB12}
\end{equation}
where $d$ is the number of spatial dimensions.

The oscillating factor $\exp(i{\bf k \cdot r})$ will kill off the contributions from large distances
unless it is compensated by a corresponding oscillation $\exp(-i{\bf k_0 \cdot r})$ in the correlation
function. Then we can write
\begin{equation}
S^{\alpha\gamma}({\bf k}) = \int d^d r \ e^{i{\bf q \cdot r}} g(r) 
\label{eqB13}
\end{equation}
where ${\bf q} = {\bf k-k_0}$, and g(r) is a smooth function. Scaling theory \cite{cardy1996,sachdev1999} then tells us that in
the vicinity of the critical point
\begin{equation}
g(r) \sim r^{-(d+z-2+\eta)}f(r/\xi)
\label{eqB14}
\end{equation}
where $\xi$ is the correlation length, and $z$ is the dynamic critical exponent. Thus when
${\bf k} = {\bf k_0}$, the `critical wavevector', we have
\begin{equation}
S^{\alpha\gamma}({\bf k_0})  =  \int d^{d} r \ r^{-(d+z-2+\eta)}f(r/\xi) 
  \sim  \xi^{2-z-\eta} \int d^{d}y \ y^{-(d+z-2+\eta)}f(y) 
\label{eqB15}
\end{equation}
where $y=r/\xi$. As the coupling $\lambda \rightarrow \lambda_c$, corresponding to a quantum phase
transition, we expect
\begin{equation}
\xi \sim |\lambda_c - \lambda|^{-\nu}
\label{eqB16}
\end{equation}
and hence
\begin{equation}
S^{\alpha\gamma}({\bf k_0}) \sim |\lambda_c - \lambda|^{-(2-z-\eta)\nu} . 
\label{eqB17}
\end{equation}

For $q = |{\bf q}|$ small but non-zero, $ q \ll 1/\xi$, we have
\bea
S^{\alpha\gamma}({\bf k}) 
 &  \sim &  \xi^{2-z-\eta} \int d^{d}y \ y^{-(d+z-2+\eta)}e^{i\xi {\bf q \cdot y}}f(y) 
 \nonumber \\
 &  \sim  & q^{-(2-z-\eta)} \int d^{d}y'\ y'^{-(d+z-2+\eta)}e^{i{\bf \hat{q} \cdot y'}}f'(y',q\xi)
\label{eqB18}
\eea
where ${\bf y'} = q\xi {\bf y}$,
so that at the critical coupling we expect $S^{\alpha\gamma}({\bf k})$ to scale like $q^{-(2-z-\eta)}$ at
small $q$.

For the 1-particle exclusive structure factor, we may paraphrase Sachdev's argument \cite{sachdev1999} as
follows. Assuming relativistic invariance of the effective field theory (i.e. $z=1$), which
applies to many though not all models, the dynamic susceptibility in the
vicinity of a quasiparticle
pole is expected to have the form
\bea
\chi^{\alpha\gamma} ({\bf k},\omega) & = & \frac{A^{\alpha\gamma}}{c^2{\bf k}^2+\Delta^2-(\omega + i\epsilon)^
2} +
\cdots
\label{eqB19}
\eea
where $\epsilon$ is a positive infinitesimal, $c$ the quasiparticle
velocity,
$\Delta$ is the quasiparticle
energy gap, and $A^{\alpha\gamma}$ is the ``quasiparticle residue".
Then the dynamic structure factor is
\bea
S^{\alpha\gamma} ({\bf k},\omega) & = & 
\frac{1}{\pi} Im\{\chi^{\alpha\gamma}({\bf k},\omega)\} 
\label{eqB20}
\eea

Let
\bea
E({\bf k}) & = & \sqrt{c^2 {\bf k}^2 + \Delta^2}
\label{eqB21}
\eea
then from (\ref{eqB19}), (\ref{eqB20}) and (\ref{eqB21}) we can write
the dynamic structure factor for the 1-particle state
\bea
S^{\alpha\gamma}_{\rm 1p} ({\bf k},\omega) & = & 
 \frac{A^{\alpha\gamma}({\bf k})}{2E({\bf k})} \delta (\omega -
E({\bf k}))
\label{eqB22}
\eea
and hence the static structure factor
\bea
S^{\alpha\gamma}_{\rm 1p}({\bf k}) & = & 
\int_0^{\infty} d\omega S^{\alpha\gamma}_{1p}({\bf k},\omega)
 = \frac{ A^{\alpha\gamma}({\bf k})}{2E({\bf k})}
\label{eqB23}
\eea
where $A^{\alpha\gamma}({\bf k})$ is the residue function,
which in general may be a function of ${\bf k}$. Note that $S({\bf k},\omega)$ at $T = 0$ vanishes for
$\omega < 0$, from equation (\ref{eqB10}).

From renormalization group theory \cite{cardy1996}, 
the scaling dimensions of these quantities are expected to be \cite{hamer2006} 
${\rm dim}[\chi^{\alpha\gamma}]  =   -2 + \eta$
and 
${\rm dim} [A^{\alpha\gamma}]  =  \eta$, 
or in other words we expect near the critical point
\bea
A^{\alpha\gamma}({\bf k}_0) & \sim & |\lambda_c-\lambda|^{\eta \nu}, 
\label{eqB24}
\eea
\bea
E({\bf k_0}) & \sim & |\lambda_c-\lambda|^{\nu},
\label{eqB24a}
\eea
and hence
\bea
S^{\alpha\gamma}_{1p}({\bf k_0}) & \sim & |\lambda_c-\lambda|^{-(1-\eta)\nu},
\label{eqB25}
\eea
just as for the total structure factor (recall here $z=1$). 
In many cases, the 1-particle contribution will dominate the structure factor, but this is not always true, as
we shall see.

These behaviours may be encapsulated in a scaling form. Assuming once again relativistic invariance
of the effective field theory near the critical point ($z = 1$), so that the quasiparticle excitation
energy is given by equation (\ref{eqB21}), and the energy gap
\bea
\Delta & = & E({\bf k_0}) \sim |\lambda_c - \lambda|^{\nu},
\label{eqB26}
\eea
then following Sachdev \cite{sachdev1999} the structure factor at low temperatures to one side of the
transition is expected to take the form
\bea
S({\bf k},\omega) = \frac{Z_S}{T^{2-\eta}}\Phi_S(\frac{cq}{T},\frac{\omega}{T},\frac{\Delta}{T})
\label{eqB27}
\eea
where $\Phi_S$ is a universal scaling function and $Z_S$ is a normalization constant depending on the
microscopic model. In the `quantum critical' regime, $\Delta/T \rightarrow 0$.

At zero temperature, we may choose $\Delta$ as the reference variable rather than $T$, and write
\bea
S({\bf k},\omega) = \frac{\tilde{Z}_S}{\Delta^{2-\eta}}\tilde{\Phi}_S(\frac{cq}{\Delta},\frac{\omega}{\Delta})
\label{eqB28}
\eea
or integrating over $\omega$,
\bea
S({\bf k}) = \frac{\tilde{Z}'_S}{\Delta^{1-\eta}}\tilde{\Phi}'_S(\frac{cq}{\Delta})
\label{eqB29}
\eea
where
\bea
\tilde{\Phi}'_S(\frac{cq}{\Delta}) = \Delta \int^{\infty}_{-\infty} d  \omega'\  \tilde{\Phi}_S (\frac{cq}{\Delta},\omega').
\label{eqB30}
\eea
If the energy gap is zero, as in the presence of Goldstone bosons, 
an energy scale can be constructed from the spin-stiffness $\rho_s$ or
the Josephson correlation length $\xi_J$ - we refer to Sachdev
\cite{sachdev1999} for details.

\section{Comparison with exact Results}
\label{sec3}

\subsection{Transverse Ising model in one space dimension}
\label{sec3a}

The transverse Ising chain model is exactly solvable, and expressions
for the energy spectrum, magnetization, etc. have been given by Pfeuty
\cite{pfeuty1970}.

Our aim is to confirm the scaling
behaviour of the structure factors for this model.
In the disordered phase, the Hamiltonian for the model can be written
as
\begin{equation}
H = \sum_i (1-\sigma^z_i) - \lambda \sum_{<ij>} \sigma^x_i \sigma^x_j
\label{eqC31}
\end{equation}
where the $\sigma^{\alpha}_i = 2S^{\alpha}_i$ are Pauli operators and the
second sum is over nearest neighbour pairs.
The critical point \cite{pfeuty1970} lies at $\lambda = 1$, and the
1-particle energy is
\begin{equation}
E(k) = 2\Lambda(k),
\label{eqC32}
\end{equation}
where
\begin{equation}
\Lambda(k) = [1+\lambda^2-2\lambda \cos (k)]^{1/2},
\label{eqC33}
\end{equation}
so that the 'critical wavevector' is $k_0 = 0$ and the energy gap is
\begin{equation}
\Delta = 2(1-\lambda).
\label{eqC34}
\end{equation}

The 1-particle exclusive structure factors have been discussed by Hamer
{\it et al.} \cite{hamer2006}.
Multiparticle expansions for correlation functions
for the quantum XY model in one space dimension
have been obtained by Vaidya and Tracy \cite{vaidya1978}. The
transverse Ising model is merely a special case of the model considered
by them
(Section 2.2 of Ref.  \cite{vaidya1978} for $t=0$, $\gamma
\rightarrow 1$, and $h=1/\lambda$). 
Hence one can obtain 
exact expressions for
the 1-particle contributions to the
correlation functions
\bea
C^{\alpha \alpha}(n) = \langle S^{\alpha}_0 S^{\alpha}_n \rangle_0
\eea
as:
\bea
C^{xx}_{\rm 1p}(n) & = &
(1-\lambda^2)^{1/4}\frac{1}{8\pi}\int^{2\pi}_0 dk
\ \frac{\cos(kn)}{\Lambda(k)}
\nonumber \\
C^{yy}_{\rm 1p}(n) & = &
(1-\lambda^2)^{1/4}\frac{1}{8\pi}\int^{2\pi}_0 dk
\ \cos(kn)\Lambda(k)
\label{C35}
\eea

Hence one finds
\bea
S^{xx}_{\rm 1p}(k) & = & \frac{(1-\lambda^2)^{1/4}}{4\Lambda (k)}
\nonumber \\
S^{yy}_{\rm 1p}(k) & = & \frac{1}{4}(1-\lambda^2)^{1/4}\Lambda (k)
\label{C36}
\eea

In the vicinity of $\lambda \rightarrow 1$, $k \rightarrow 0$, equation
(\ref{eqC32}) reduces to
\bea
E(k) \rightarrow \Delta f(cq/\Delta)
\label{eqC36a}
\eea
where
\bea
f(x) = \sqrt(1+x^2)
\label{eqC37}
\eea
with $c=2$, which is the expected relativistic form. The 1-particle
structure factor $S^{xx}_{1p}$ reduces to
\bea
S^{xx}_{1p}(k) \rightarrow \frac{1}{2} \Delta^{-3/4}
\tilde{\Phi}^{'xx}_{1p}(cq/\Delta)
\label{eqC38}
\eea
which has the expected scaling form (c.f. equation(\ref{eqB29})), with
$d=1, z=1, \eta = 1/4, \nu=1$, the transverse Ising model values, and
\bea
\tilde{\Phi}^{'xx}_{1p}(cq/\Delta) = 1/f(cq/\Delta).
\label{eqC39}
\eea
The other transverse structure factor
\bea
S^{yy}_{1p}(k) \rightarrow \frac{1}{8} \Delta^{5/4}
\tilde{\Phi}^{'yy}_{1p}(cq/\Delta).
\label{eqC40}
\eea
where
\bea
\tilde{\Phi}^{'yy}_{1p}(cq/\Delta) & = & f(cq/\Delta).
\label{eqC40a}
\eea
Note that whereas $S^{xx}_{\rm 1p}(k)$ diverges as $\{\lambda \to 1, k=0\}$,
$S^{yy}_{\rm 1p}(k)$ does not, and has a sub-leading critical index, two
powers of $\Delta$ smaller than $S^{xx}_{1p}$. It appears that $S^{yy}$ decouples from the one-particle
state at the critical point.

The quasiparticle residue for the dominant spectral weight $S^{xx}$ at $k=0$ is
\bea
A(k) & = & (1-\lambda^2)^{1/4} \sim [2(1-\lambda)]^{1/4}, \  \lambda
\rightarrow 1,
\label{eqC40a}
\eea
in agreement with Sachdev's result \cite{sachdev1999}, after one takes into
account differing normalization factors in our definitions.
Note that in this case $A(k)$ is independent of $k$.

We may deduce the scaling form of the full 1-particle structure
function in the vicinity of the critical point:
\bea
S^{xx}_{1p}(k,\omega) & = &
\frac{\Delta^{-7/4}}{2f(cq/\Delta)}\delta(\omega/\Delta
-f(cq/\Delta))
\label{C41}
\eea
whence the scaling function for the dominant component  may be taken as
\bea
\tilde{\Phi}^{xx}_{1p}(cq/\Delta,\omega\Delta) & = &
\tilde{\Phi}'^{xx}_{1p}(cq/\Delta)\delta(\omega/\Delta
-f(cq/\Delta))
\label{eqC42}
\eea
with normalization factor $\tilde{Z}^{xx}_{1p} = 1/2$.
 These are the simplest possible
free-particle forms, save only the renormalization of the residue
function with coupling.

\section{Comparison with Numerical Results}
\label{sec4}

\subsection{The Transverse Ising model in higher dimensions}
\label{sec3b}

\begin{figure}[h]
\begin{minipage}{18pc}
\includegraphics[width=18pc]{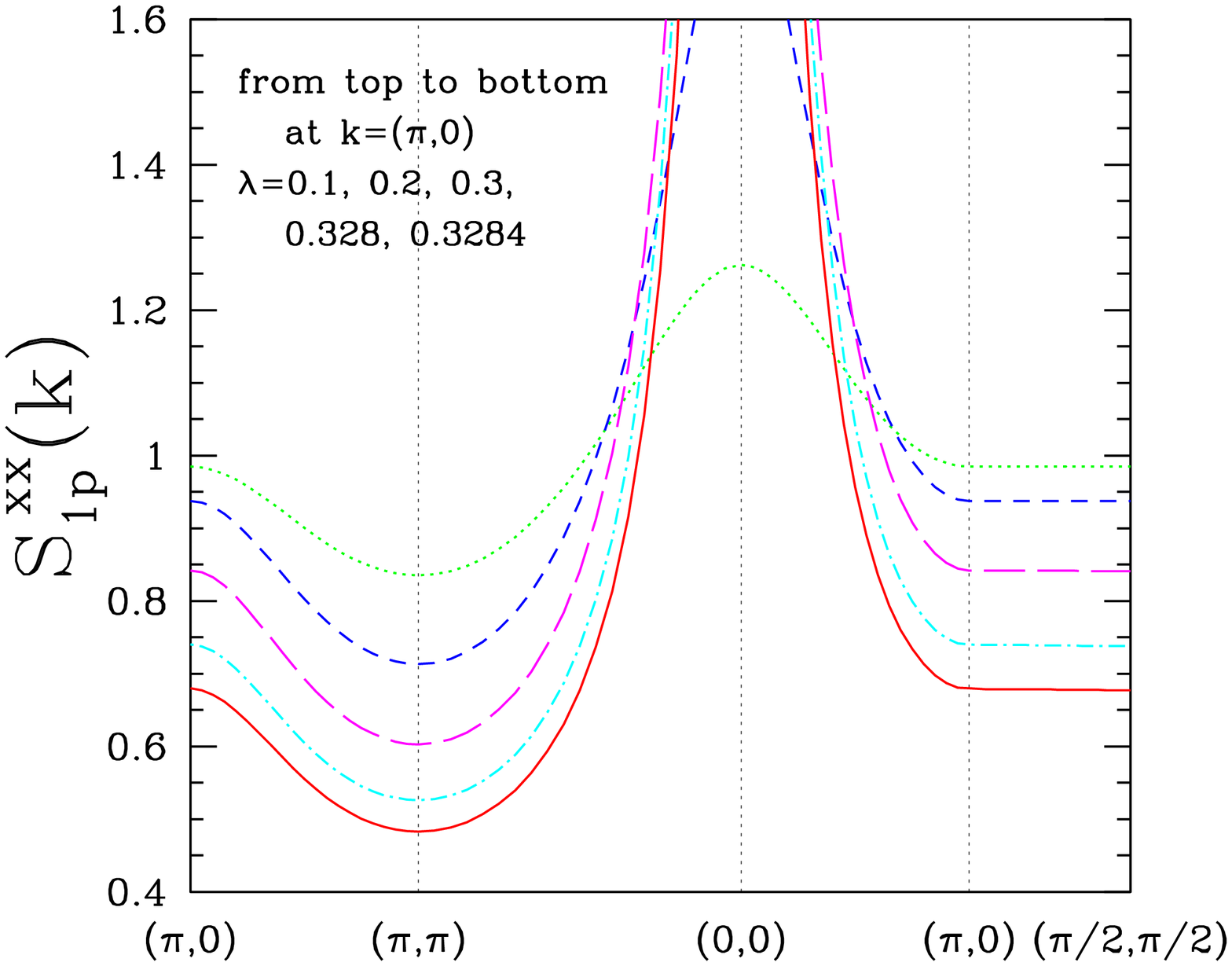}
 \caption{ \label{fig1}
(Color online) $S_{\rm 1p}^{xx} ( {\bf k} )$ along high-symmetry cuts through the Brillouin
zone for the transverse Ising model  with coupling $\lambda=0.1$, 0.2, 0.3, 0.328, 0.3284 on
the square lattice.
 (From ref. \cite{hamer2006}).
}
\end{minipage}\hspace{2pc}%
\begin{minipage}{22pc}
\includegraphics*[width=18pc]{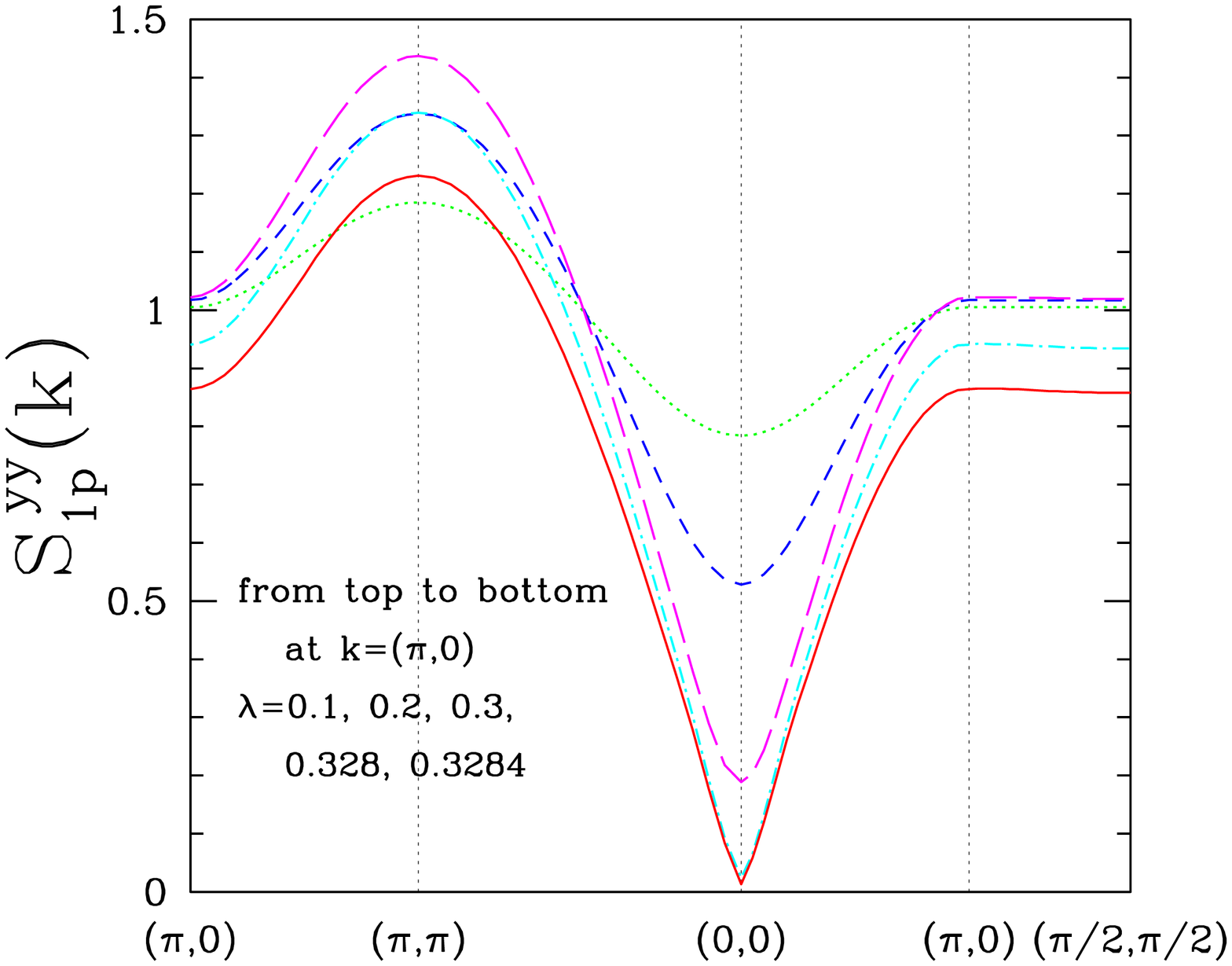}
 \caption{\label{fig2}
(Color online) $S_{\rm 1p}^{yy} ( {\bf k} )$ along high-symmetry cuts through the Brillouin
zone for the transverse Ising model with coupling $\lambda=0.1$, 0.2, 0.3, 0.328, 0.3284 on
the square lattice.
 (From ref. \cite{hamer2006}).
}
\end{minipage}
\end{figure}

\begin{figure}[h]
\begin{minipage}{18pc}
\includegraphics[width=18pc]{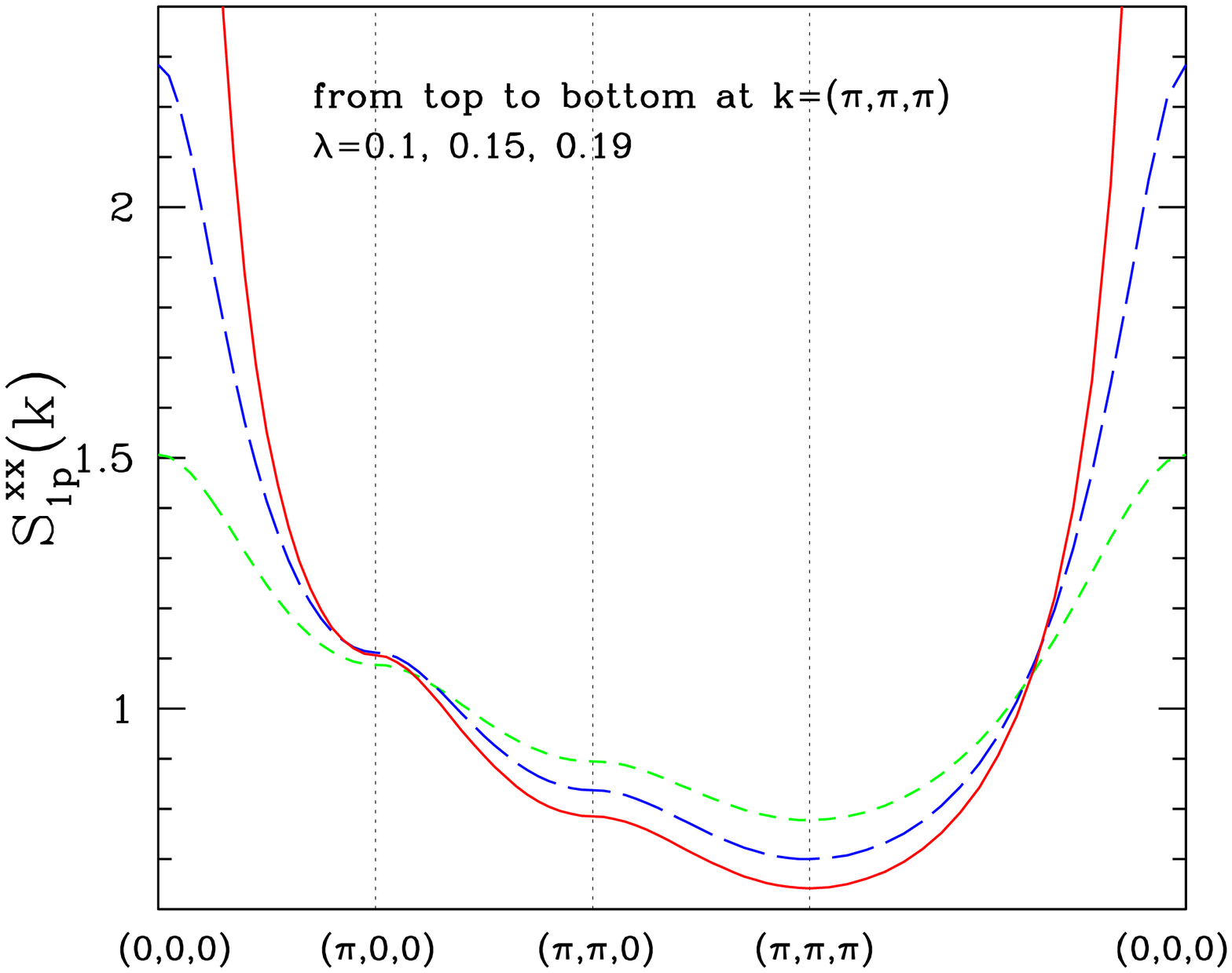}
\caption{\label{fig3}
(Color online) $S_{\rm 1p}^{xx} ( {\bf k} )$ along high-symmetry cuts through the Brillouin
zone for the transverse Ising model with coupling $\lambda=0.1$, 0.15 and 0.19 on the simple cubic lattice.
 (From ref. \cite{hamer2006}).
}
\end{minipage}\hspace{2pc}%
\begin{minipage}{22pc}
\includegraphics*[width=18pc]{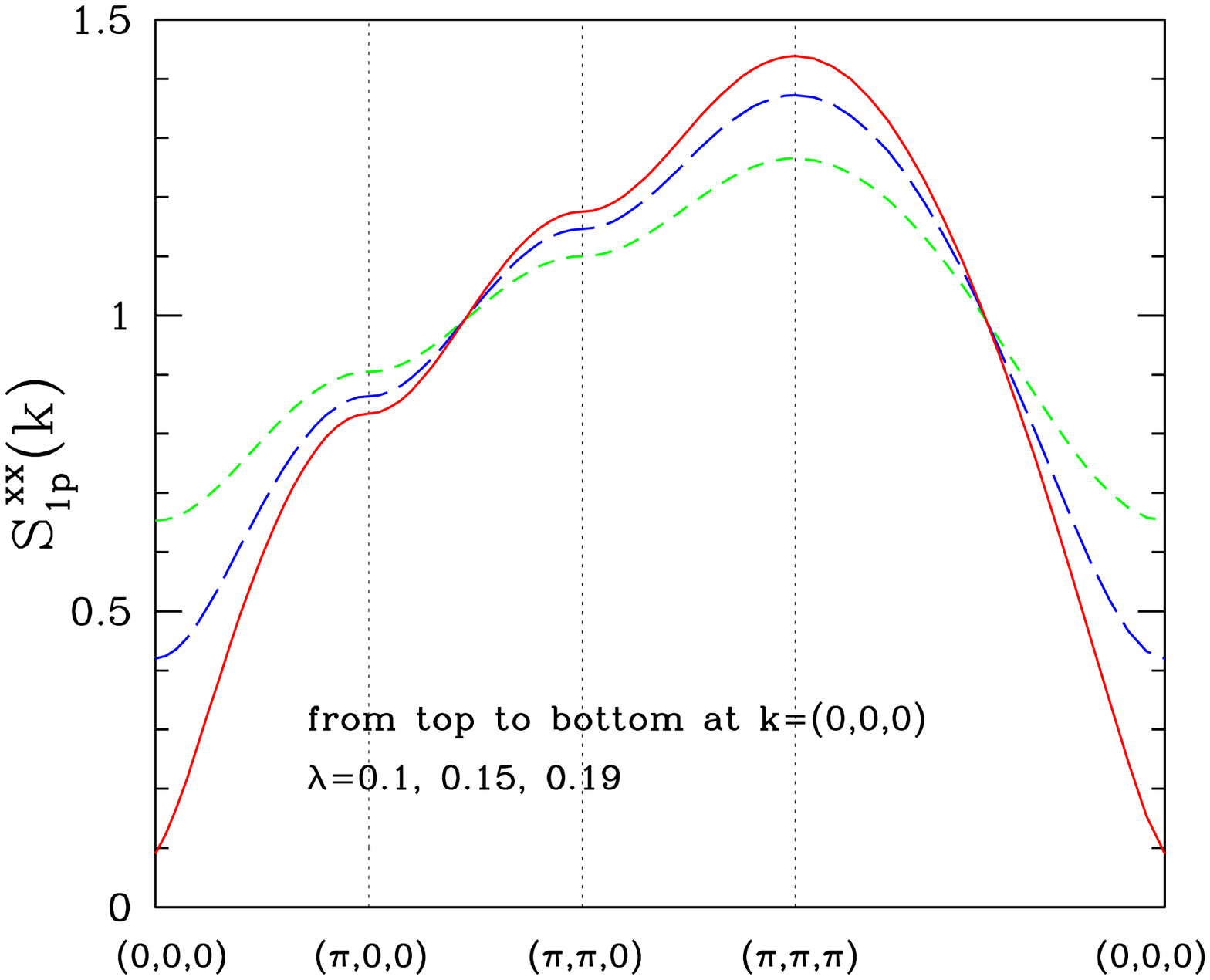}
\caption{\label{fig4}
(Color online) $S_{\rm 1p}^{yy} ( {\bf k} )$ along high-symmetry cuts through the Brillouin
zone for the transverse Ising model with coupling $\lambda=0.1$, 0.5 and 0.19 on the simple cubic lattice.
 (From ref. \cite{hamer2006}).
}
\end{minipage}
\end{figure}

The behaviour of the transverse Ising model in higher dimensions is qualitatively
similar. The 1-particle structure factors for the transverse Ising model
on the triangular, square, and cubic lattices have also been calculated by
Hamer {\it et al.} \cite{hamer2006}, using high-order series expansions.
Some sample results for the square and cubic lattices are shown in
Figures \ref{fig1}-\ref{fig4}. 

For the square lattice, the critical point is estimated 
\cite{hamer2000} to lie at $\lambda = 0.32841(2)$,
and the critical exponents are expected to be the same as those of the
classical 3D Ising model, namely $\eta = 0.0364(5)$, $\nu = 0.6301(4)$, from various estimates
\cite{pelissetto2002}.
The results for $S_{\rm 1p}^{xx}$ and $S_{\rm 1p}^{yy}$  along high-symmetry cuts through the Brillouin
zone for the system with couplings $\lambda=0.1$, 0.2, 0.3, 0.328 and 0.3284
 are given in Figures \ref{fig1} and \ref{fig2}.
 The results of a standard Dlog Pad{\' e} analysis \cite{hamer2006} of the series for 
$S^{xx}_{\rm 1p}({\bf k})$ at
 ${\bf k}=(0,0)$ and ${\bf k}=(\pi/2,\pi/2)$
 at ${\bf k}=(0,0)$, where the energy gap vanishes, give estimates  $\lambda_c= 0.3284(4)$ with
 exponent $-0.605(5)$,
 compared to the expected exponent $\nu(\eta-1) = -0.607$.
 At momentum ${\bf k}=(\pi/2,\pi/2)$, where the energy gap remains finite,
 we find $\lambda_c =0.34(3)$ with exponent $ 0.04(2)$ compared to the
 expected value $\nu\eta = +0.0229$. 
For $S_{\rm 1p}^{yy}$, the  estimate for the critical index
 is very close to the value
 $\nu(\eta+1)=0.65$.

In Figures \ref{fig1} and \ref{fig2} for $\lambda=0.328$ and  0.3284, we have biased the critical point to
$\lambda_c=0.32841$
 with critical index $\nu\eta = +0.0229$ in our analysis. 
 We can see from these figures, that even for
 $\lambda=0.3284$ which is very close to the critical point, $S_{\rm 1p}^{xx}$ and $S_{\rm 1p}^{yy}$
are  still far from zero. This reflects the tiny value of the exponent $\eta\nu$, which implies
a precipitous drop to zero just before the critical point.

Figures \ref{fig3} and \ref{fig4} show similar graphs for the simple
cubic lattice. In this case, 
the critical point has been obtained previously \cite{zheng1994} as
$\lambda_c = 0.19406(6)$,
and the critical exponents are expected to lie
in the universality
class of the 4D classical Ising model, where we expect the mean field exponents
 $\eta =0$ , $\nu =1/2$, modulo logarithmic corrections \cite{sachdev1999}.

The analysis of $S^{xx}_{\rm 1p}({\bf k})$ at ${\bf k}=(0,0,0)$,
where the energy gap vanishes, gives $\lambda_c= 0.19406(8)$ with
exponent $-0.54(1)$, while for  $S^{yy}_{\rm 1p}({\bf k})$ at ${\bf k}=(0,0,0)$,
the estimate of the critical point is $\lambda_c= 0.194(4)$ with
exponent $0.55(3)$. Away from ${\bf k}=(0,0,0)$,
 where the energy gap remains finite,
we find $\lambda_c =0.22(3)$ with exponent $ 0.03(2)$ for both $S^{xx}_{\rm 1p}({\bf k})$
and $S^{yy}_{\rm 1p}({\bf k})$. Allowing for
logarithmic corrections, these estimates agree reasonably well with the expected values.

In all cases, we see that the dominant structure factor $S^{xx}_{1p}$ at the critical wavevector diverges
at the critical coupling 
with exponent $-\nu(1-\eta)$, while $S^{yy}_{1p}$ vanishes with exponent consistent with $\nu(1+\eta)$.
Away from the critical wavevector, the structure factors both vanish at the critical coupling 
with a small exponent consistent
with $\nu\eta$.

\subsection{The Alternating Heisenberg Chain}
\label{sec3c}

Schmidt and Uhrig \cite{schmidt2003} and Hamer {\it et al.} \cite{hamer2003} have investigated
the spectral weights of
the  alternating Heisenberg chain, which can be described
by the following Hamiltonian
\begin{equation}
H =  \sum_{i}  \left( {\bf S}_{2i}\cdot {\bf S}_{2i+1} +
 \lambda {\bf S}_{2i-1}\cdot {\bf S}_{2i} \right)
\label{eqD43}
\end{equation}
where the ${\bf S}_i$ are  spin-$\frac{1}{2}$ operators at site $i$,
and $\lambda$ is the alternating coupling. Here we assume that the distance
between neighboring spins are all equal and the distance
between two successive dimers is $d$.

There is a considerable literature on this model, which has been reviewed
by Barnes et al. \cite{barnes1999}.
At $\lambda = 0$, the system consists of a chain of decoupled dimers,
and in the ground state each dimer is in a singlet
state. Excited states are made up from the three triplet excited
states on each dimer, with a finite energy gap between the singlet ground state
and the triplet excited states. This scenario is
believed \cite{duffy1968,bonner1982,jiang2001} to hold
right up to the uniform limit $\lambda = 1$, which corresponds to a
critical point. At $\lambda = 1$, we regain the uniform Heisenberg chain,
which is gapless.

Several theoretical
papers \cite{den1979,uhrig1999,sorenson1998,tsvelik1997} have discussed the
approach to the uniform limit.
Analytic studies of the critical behaviour near $\lambda=1$ \cite{den1979}
have related the alternating chain to the 4-state Potts model, and
indicate that the ground-state energy per site $\epsilon_0(\lambda)$, and
the energy gap $\Delta(\lambda)$ should behave as
\bea
\epsilon_0 (\lambda) - \epsilon_0 (1) &\sim& \delta^{4/3}/|\ln (\delta/\delta_0)| 
\\
\Delta (\lambda) &\sim& \delta^{2/3}/\sqrt{|\ln (\delta/\delta_0 )|} 
\label{eqD44}
\eea
as $\lambda\to 1$, where $\delta=(1-\lambda)/(1+\lambda)$.
This corresponds to critical exponents $\alpha=2/3$, $\nu = 2/3$.
The logarithmic terms in (\ref{eqC42}) are due to the existence of a marginal variable
in the model.

For the uniform chain $\lambda=1$, and near $kd\to 2\pi$, Affleck 
\cite{affleck1998}
has obtained expressions for the correlation functions in the model,
including logarithmic corrections, which correspond to an exponent $\eta
= 1$:
\bea
G^z(r) = G^x(r) \rightarrow \frac{1}{(2\pi)^{3/2}}\frac{(\ln
r)^{1/2}}{r}.
\label{eqD44a}
\eea
Fourier transforming, one obtains
the asymptotic form for $S^{-+}(kd)$ as
\bea
S(kd) \equiv S^{-+}(kd) = \frac{8}{3 (2\pi)^{3/2}} |\ln (\pi - kd/2) |^{3/2}
\label{eqD44b}
\eea
Note that in this case $(1-\eta)\nu = 0$, so there is no power-law
divergence in the structure factor, but rather a logarithmic one.

This implies that for $kd=2\pi$ and  as $\lambda\to 1$,
the asymptotic form for
$S(2\pi)$ diverges as
\bea
S(2\pi) \propto [- \ln (1-\lambda)]^{3/2} \quad \lambda \to 1 
\label{eqD45}
\eea
For $0<kd<2\pi$, one expects $S$ to be finite for any $\lambda$.

\begin{figure}[h]
\begin{minipage}{18pc}
\includegraphics[width=18pc]{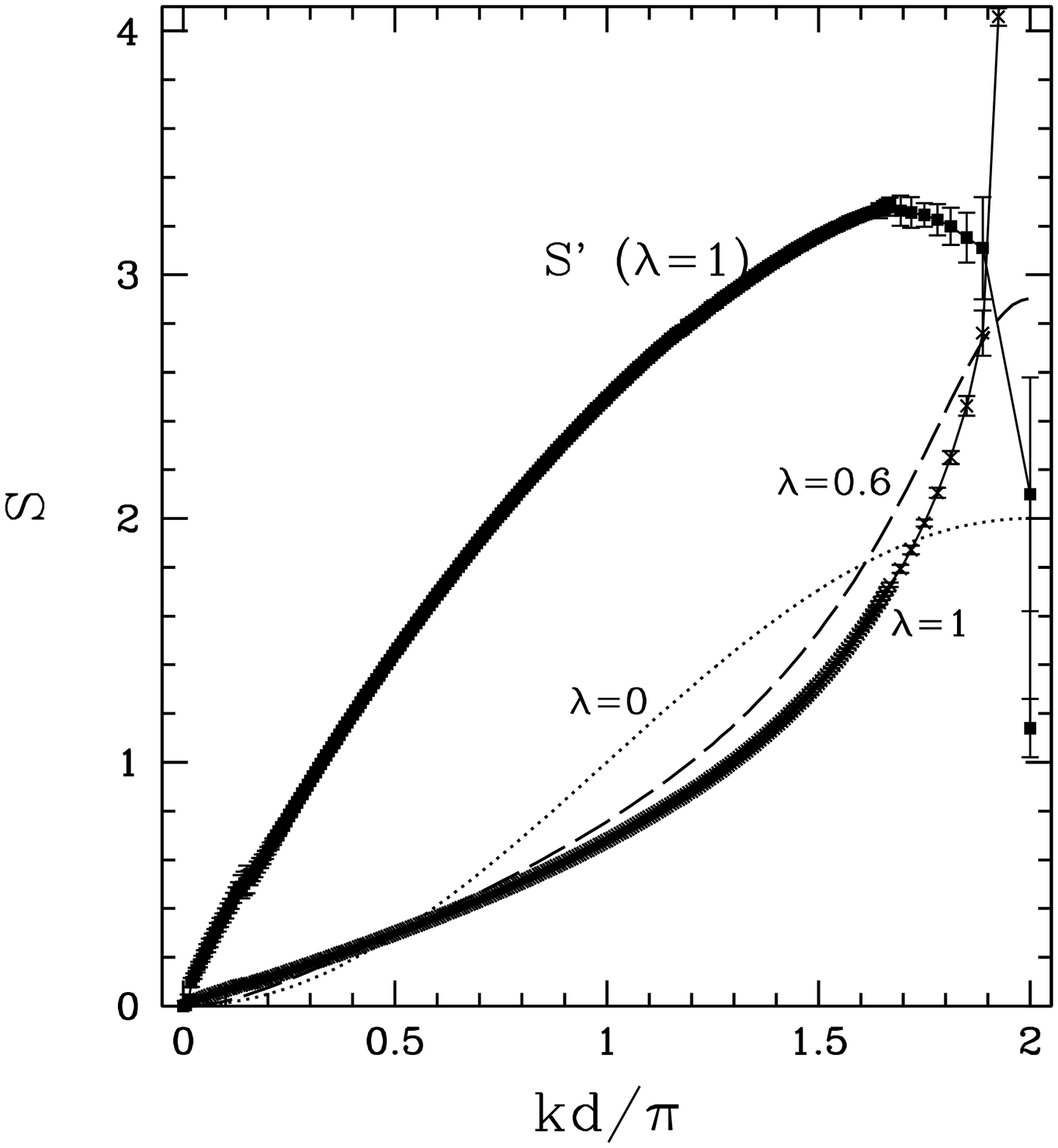}
\caption{\label{fig5}
         The integrated structure factor $S$ versus $k$ for
the alternating Heisenberg chain with
         $\lambda=0$ (dotted line), 0.6 (dashed line),  1 (crosses).
         Also shown is the quantity
          $S'\equiv 6 S [- 2\pi\ln(1-\frac{k}{2\pi})/k]^{-3/2}$
           for $\lambda=1$ (squares).
(From ref. \cite{hamer2003}).}
\end{minipage}\hspace{2pc}%
\begin{minipage}{22pc}
\includegraphics*[width=18pc]{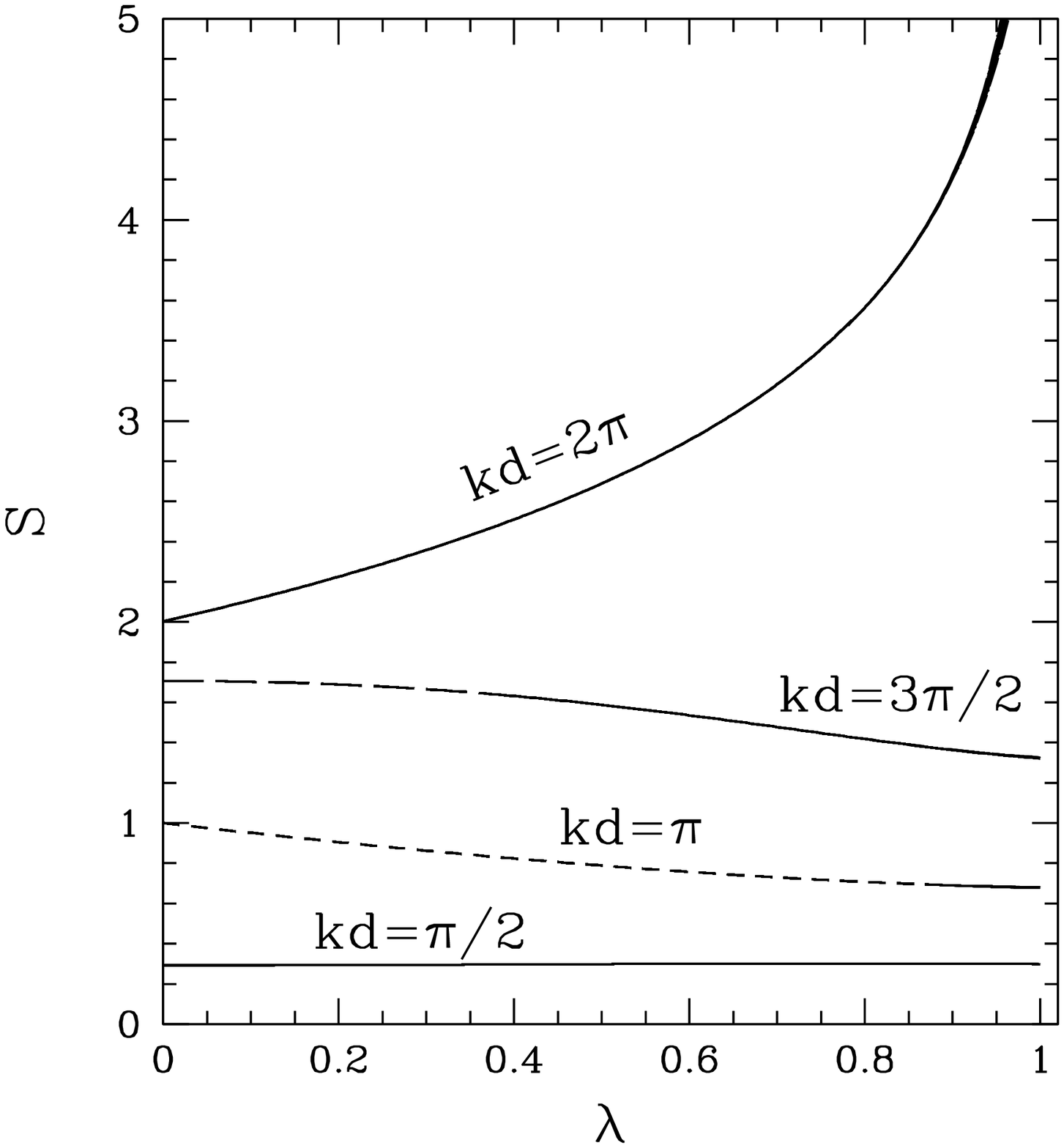}
\caption{\label{fig6}
         The integrated structure factor $S$ versus $\lambda$ for
the alternating Heisenberg chain with
         $kd=\pi/2$, $\pi$, $3\pi/2$ and $2\pi$.
(From ref. \cite{hamer2003}).}
\end{minipage}
\end{figure}

The results obtained by Hamer {\it et al.} \cite{hamer2003} for $S$ versus momentum $k$ for
$\lambda=0$, 0.6, and 1 are shown in Fig. \ref{fig5}.
Note that $\int_0^{2\pi} S(k) dk= 2 \pi$ (here we set $d=1$),
independent of 
$\lambda$, so
the area under each curve is the same.
Also shown in the figure are the results for
$S'\equiv 6 S [- 2\pi\ln(1-\frac{k}{2\pi})/k]^{-3/2}$ at
$\lambda=1$.
The results appear reasonably consistent with the expected behaviour.

For fixed values of $k$, Fig. \ref{fig6} shows the
integrated structure factor $S$ versus $\lambda$, where for each value
of $k$, about 20 different integrated differential approximants
to the series are shown. We can see that the results converge
very well out to $\lambda=1$. The logarithmic divergence as $\lambda \to 1$
for the case $kd=2\pi$ is clearly evident.

For $0<kd<2\pi$, an analysis of the series for the 1-particle structure
factor $S_{\rm 1p} \equiv S^{-+}_{1p}$ using
Dlog Pad\'e approximants by Schmidt and Uhrig \cite{schmidt2003}
appeared to show that it vanishes with a behavior close to
 $(1-\lambda)^{1/3}$. Since $S$ remains finite, one would thus expect
that $S_{\rm 1p}/S$ vanishes like $(1-\lambda)^{1/3}$.
This agrees with a heuristic argument \cite{schmidt2003}
that the 1-particle spectral weight should vanish like
$\sqrt{\Delta}$, i.e.  like $\delta^{1/3}/|\ln (\delta/\delta_0)|^{1/4}$,
where $\delta =(1-\lambda)/(1+\lambda)$.
It {\it disagrees}, however, with 
what one might expect from the transverse Ising model example, that the one-particle residue
should vanish with exponent $\eta\nu = 2/3$ at all wavevectors, leading to
a behaviour $(1-\lambda)^{2/3}$. 
It is possible that a logarithmic
correction term may again be disguising the true power-law behaviour; or alternatively, the
power-law behaviour of the renormalized one-particle residue function might indeed be different
away from the critical wavevector.
It would
be useful to have some further analytical guidance in this case. 

\begin{figure}[h]
\begin{minipage}{18pc}
\includegraphics[width=18pc]{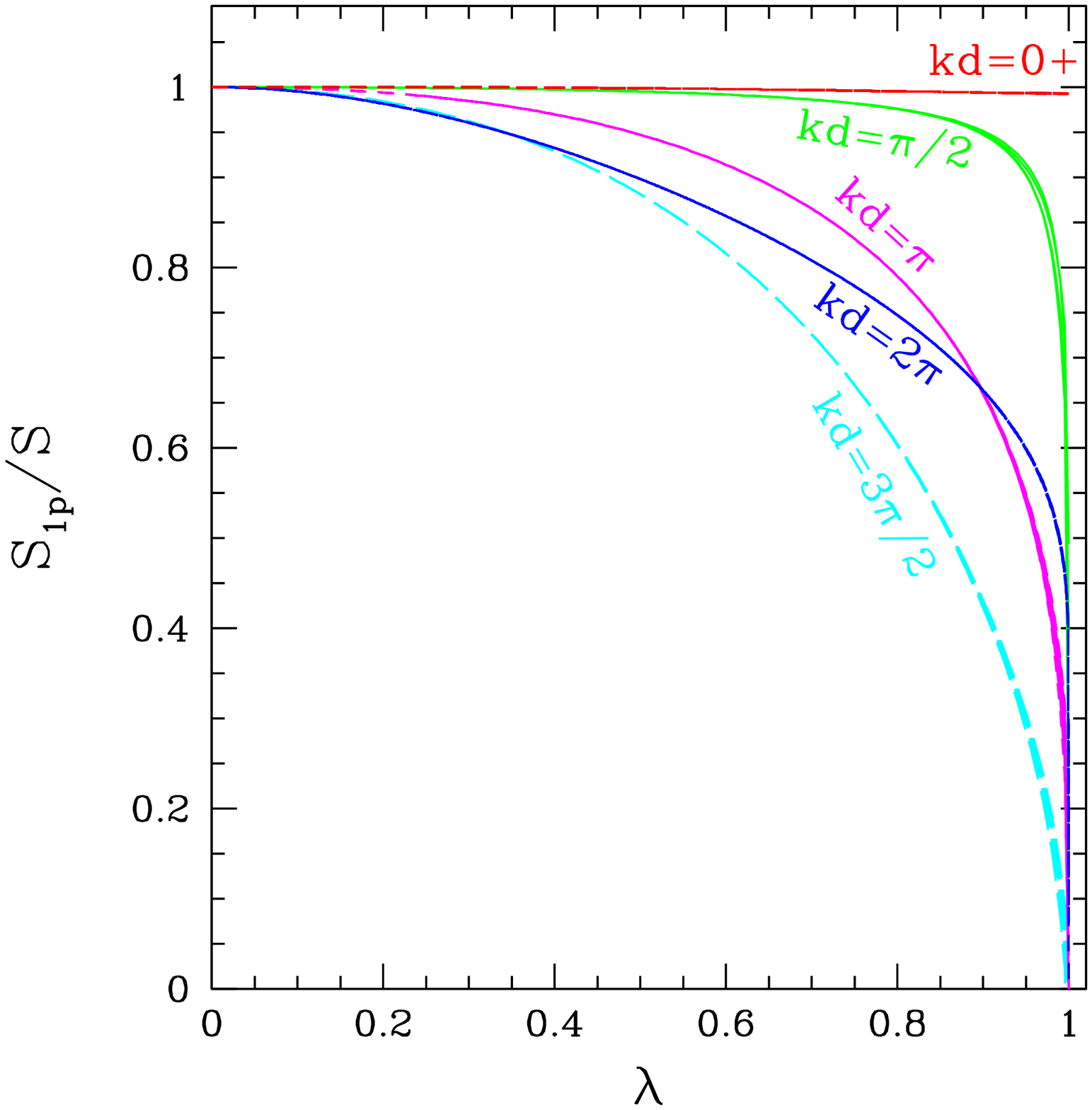}
\caption{\label{fig7}
         The relative 1-particle weight $S_{\rm 1p}/S$ versus $\lambda$ for
the alternating Heisenberg chain with
         $kd=0+$, $\pi/2$, $\pi$, $3\pi/2$ and $2\pi$.
(From ref. \cite{hamer2003}).
}
\end{minipage}\hspace{2pc}%
\begin{minipage}{22pc}
\includegraphics*[width=18pc]{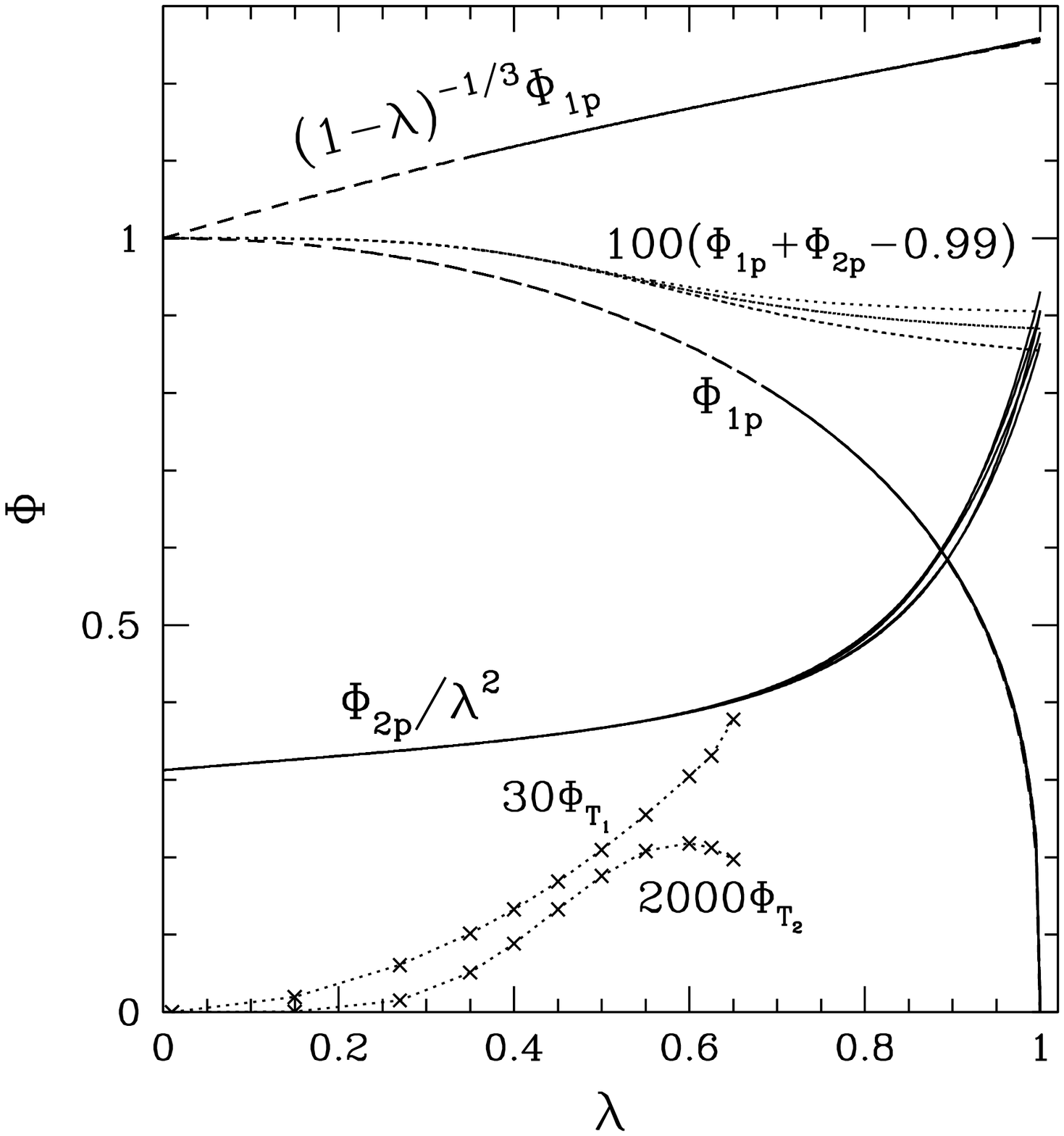}
\caption{\label{fig8}
         The auto correlation functions versus $\lambda$
of the bilayer Heisenberg model
         for the 1-particle state ($\Phi_{\rm 1p}$),
         2-particle states  ($\Phi_{\rm 2p}$), and two particle bound states
         $T_1$ and $T_2$.
(From ref. \cite{collins2008}).}
\end{minipage}
\end{figure}

Fig. \ref{fig7} shows numerical values from Hamer {\it et al.}
\cite{hamer2003} for the
relative 1-particle weight $S_{\rm 1p}/S$ versus $\lambda$
at selected values of $kd$.
It can be seen that for any non-zero value of
$k$, $S_{\rm 1p}/S$ decreases abruptly to zero as $\lambda\to 1$. Only at  $kd= 0+$,
does $S_{\rm 1p}/S$ remain finite (about 0.993) in the limit $\lambda=1$;
but by then $S$ has itself decreased to zero.

Finally, we discuss the results for the spin auto-correlation functions,
defined as 
\bea
\Phi(\omega) = \frac{1}{2\pi} \int^{\infty}_{-\infty} dk
S^{-+}(k,\omega)
\label{eqD46}
\eea

Schmidt and Uhrig \cite{schmidt2003} argued that the critical behaviour for
the total auto correlation function (summed over $\omega$) of
the 1-particle state $\Phi_{\rm 1p}$  should be
\bea
\Phi_{\rm 1p} &\propto & (1-\lambda)^{1/3}  \label{C_Phi1p}  
   \label{eqD47}
\eea
modulo logarithms, as for the structure factors.

Figure \ref{fig8} shows various auto-correlation functions versus $\lambda$,
reproduced from Hamer {\it et al.}.
One can see that $\Phi_{\rm 1p}$ vanishes at the limit
$\lambda=1$, while $(1-\lambda)^{-1/3} \Phi_{\rm 1p}$ increases almost linearly
as $\lambda$ increases.
The curve for $(\Phi_{\rm 1p} +\Phi_{\rm 2p}) $,
if we {\it assume} it is non-singular at $\lambda=1$ (i.e. the
singularities in $\Phi_{\rm 1p}$ and $\Phi_{\rm 2p}$ cancel exactly),
runs almost flat with $\lambda$ once we neglect unphysical and defective
approximants: that would indicate that the 2-particle sector accounts for
about 99.8\% of the weight, even at $\lambda=1$,
which agrees almost exactly with the conclusions of Schmidt and
Uhrig \cite{schmidt2003}. Remarkably, this is much higher than the
fraction of 73\% for the
two-spinon continuum at $\lambda=1$ calculated by Karbach {\it et al.} \cite{karbach1997}
from the exact solution.
Also shown in Fig. \ref{fig8} is the direct
extrapolation of the 2-particle auto-correlation $\Phi_{\rm 2p}$ using
integrated differential approximants. These extrapolations assume that
there is {\it no}
singularity in $\Phi_{\rm 2p}$ at $\lambda=1$, and the results
give a somewhat smaller value of about 0.9  at $\lambda=1$.

Overall, then, the 1-particle energy gap and spectral weight at general momenta appear to
vanish as $\lambda\to 1$, following the behaviour predicted by Cross and Fisher
\cite{den1979},
and already confirmed numerically by Singh and Zheng \cite{singh1999}.
However, the 2-triplet spectral weight remains finite in the uniform limit
and, in fact, appears to form the major part of the
total spectral weight. Schmidt and
Uhrig \cite{schmidt2003} already pointed out that indeed the 2-triplet states carry a larger portion
of the total spectral weight than the 2-spinon states, calculated by Karbach
{\it et al.} \cite{karbach1997}. 
This argues that a description in terms of triplons remains equally
valid with a description in terms of spinons for the uniform chain.

\subsection{Heisenberg Bilayer Model}
\label{sec3a}

As our final example, we consider the Heisenberg bilayer antiferromagnet
on the
square lattice, with Hamiltonian
\begin{equation}
H =  J_1 \sum_{l \ = \ 1,2}
\sum_{<i,j>} {\bf
S_{l i} \cdot S_{l j}}
+  J_2 \sum_{ i } {\bf S_{1i} \cdot S_{2i}}
\label{eq1}
\end{equation}
where $l = 1,2$ labels the two planes of the bilayer. The physics of the
system then depends on the coupling ratio $\lambda = J_1/J_2$. At
$\lambda=0$, 
the
ground state consists simply of $S = 0$ dimers on each bond between the
two
layers, and excitations are composed of $S = 1$ `triplon' states
\cite{schmidt2003}
on one or
more bonds. At large $\lambda$, where the $J_1$ interaction is dominant,
the ground
state will be a standard N{\' e}el state, with $S = 1$ `magnon'
excitations.
At some intermediate critical value $\lambda_c$, a phase transition will
occur
between these two phases. It is believed that this transition is of
second order,
and is accompanied by a Bose-Einstein condensation of
triplons/magnons in the ground state.

\begin{figure}[h]
\begin{minipage}{18pc}
\includegraphics[width=18pc]{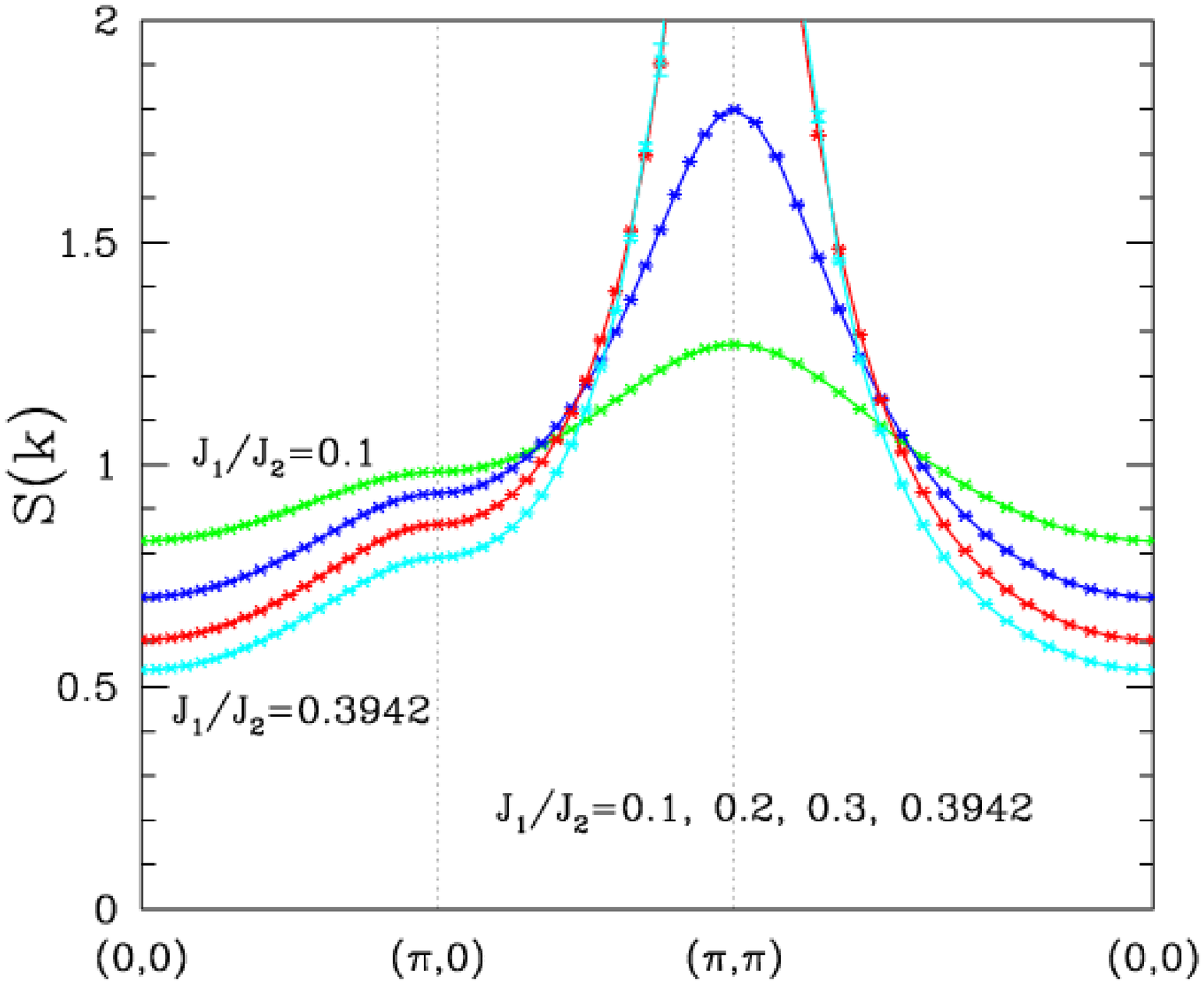}
\caption{\label{fig9}
The total static structure factor $S({\bf k})$ in the bilayer Heisenberg model 
as a function of ${\bf
k}$ at various couplings $\lambda = J_1/J_2$. 
(From ref. \cite{collins2008}).
}
\end{minipage}\hspace{2pc}%
\begin{minipage}{22pc}
\includegraphics*[width=18pc]{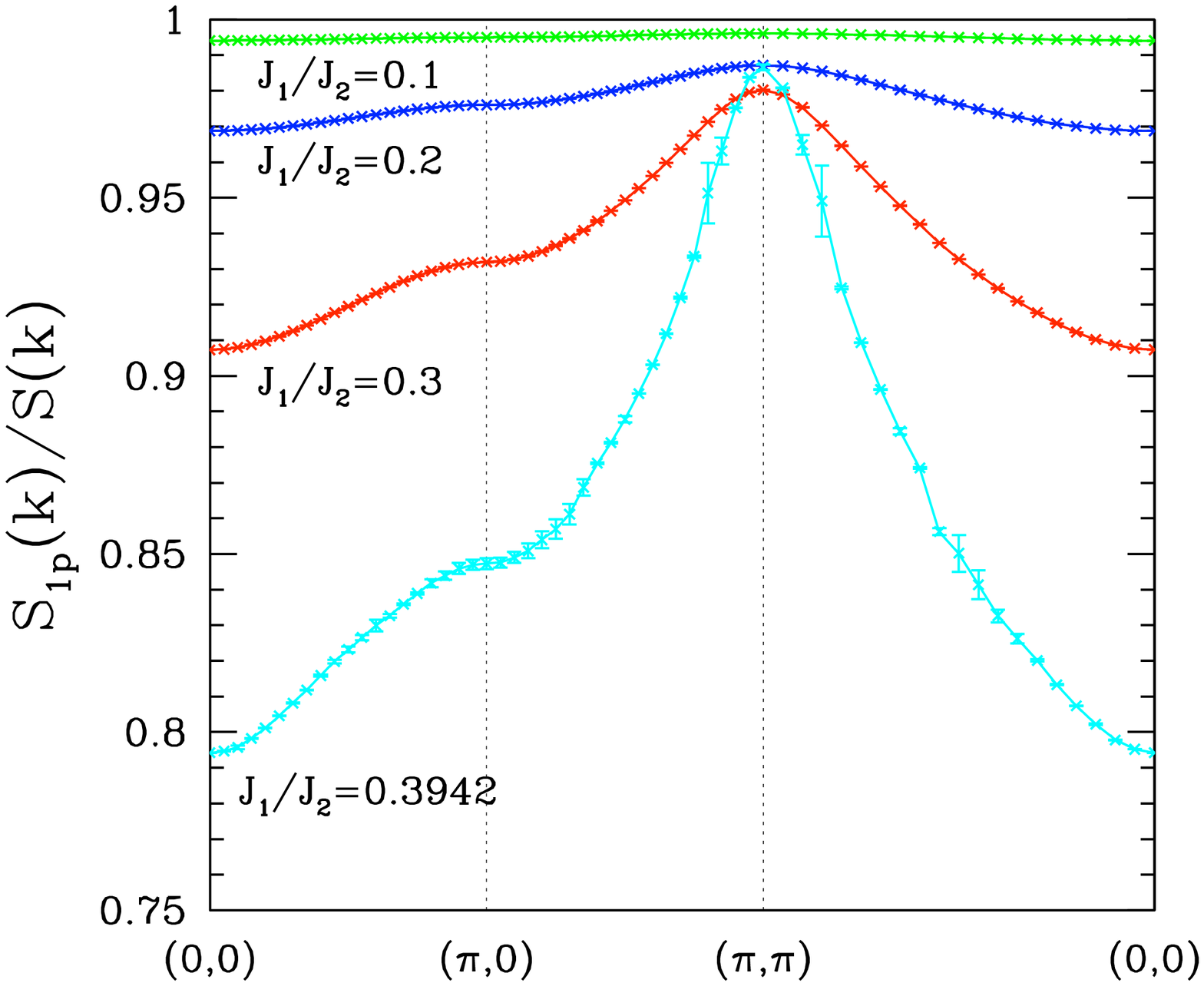}
\caption{\label{fig10}
The ratio $S_{1p}({\bf k})/S({\bf k})$ of the 1-particle static
structure factor to the total static structure factor as a function of
${\bf k}$ in the bilayer Heisenberg model, for various couplings $\lambda = J_1/J_2$.
(From ref. \cite{collins2008}). }
\end{minipage}
\end{figure}

Figures \ref{fig9} and \ref{fig10} show some series results for
structure factors in the dimerized phase, calculated by Collins and
Hamer \cite{collins2008}. Figure \ref{fig9}
shows the total static transverse structure factor $S({\bf k}) \equiv S^{+-}({\bf
k})$ as a function of ${\bf k}$ at various couplings $\lambda =
J_1/J_2$.
All results are for $k_z = \pi$, probing intermediate states antisymmetric between the planes, 
and we only refer to
${\bf k} = (k_x,k_y)$ hereafter.
 
 The dominant feature is a large peak at the N{\' e}el point
${\bf k} = (\pi,\pi)$, which appears to become divergent as $\lambda
\rightarrow \lambda_c$, as we would expect. 
 Figure \ref{fig10} shows the ratio of the 1-particle
structure factor $S_{1p}({\bf k})$ to the total $S({\bf k})$ as a
function of ${\bf k}$. The 1-particle contribution generally remains
the dominant part of the total, particularly near the N{\' e}el point.

Let us now compare these results with theoretical expectations. From
scaling theory (Sec. \ref{sec2}),
both the 1-particle structure
factor and the total structure factor in the vicinity of the critical point 
should scale like
$(\lambda_c-\lambda)^{(\eta-1)\nu}$,
at the critical (N{\' e}el) momentum.
We expect this transition to belong to the universality class of the
O(3) model in 3 dimensions, which has critical exponents
\cite{guida1998}
$\nu = 0.707(4)$, $\eta = 0.036(3)$, hence we expect $(\eta-1)\nu =
-0.682(5)$, which is quite compatible with the numerical estimates.

How does $S_{1p}$ behave at the critical coupling
away from the N{\' e}el momentum?
Here the behaviour is quite different from the previous models. 
The ratio $S_{1p}/S$ decreases smoothly towards the critical coupling, 
and shows no sign of vanishing there.
In fact the 1-particle structure factor remains dominant everywhere, remaing at
80\% of the total or more.
Thus it appears that in this case the renormalized residue function does not vanish
at $\lambda_c$, except at the N{\' e}el momentum.

\section{Summary and Conclusions}
\label{sec5}

This paper consists largely of a review of the behaviour of structure
factors near a quantum phase transition, at temperature $T = 0$. We have
focused here on quantum spin models, but the conclusions should apply
more generally.

Section \ref{sec2} reviewed current theory on the subject, drawn largely
from Sachdev \cite{sachdev1999}. The generic scaling
behaviour of both the total structure factor and the 1-particle
exclusive structure factor is predicted to be the same, determined by
the critical exponents $\eta$ and $\nu$.

We then reviewed calculations of the structure factors for some specific
quantum spin models. For the transverse Ising model in one dimension,
exact results can be obtained \cite{hamer2006}; while for the transverse Ising
model in higher dimensions \cite{hamer2006}, the alternating Heisenberg chain
\cite{schmidt2003,hamer2003}, and the bilayer Heisenberg model \cite{collins2008}, we have used some
numerical results obtained from series expansions to high orders. For
the most part, the results conform to theoretical expectations.

Some significant differences have been noted, however, in the detailed
behaviour of these models, particularly as regards the 1-particle
structure factor. In the transverse Ising model 
the 1-particle residue vanishes at the critical point
for all wavevectors, and so the 1-particle contribution to the total
structure factor becomes negligible. 
For the solvable case of the one-dimensional chain, the residue is actually independent of
wavevector.

For the alternating chain, the one-particle residue again vanishes at the critical point,
and
it is the 2-particle `triplon' state which appears to become
dominant at the phase transition \cite{schmidt2003,hamer2003}. 
But the residue appears to vanish with a different exponent depending on the wavevector, namely
2/3 at the critical wavevector and 1/3 away from it, which seems peculiar. It could be that the
true exponent is disguised by logarithmic corrections, or perhaps the renormalized residue
function does indeed behave differently at different wavevectors, and vanishes with a subdominant
exponent away from the critical wavevector. Further analysis is needed
here.

For the bilayer Heisenberg
model, on the other hand, the renormalized 1-particle residue vanishes
at the critical wavevector only, and the 1-particle state remains
dominant at the critical point. This is presumably the more typical
pattern of behaviour.

\ack
This work is supported by a grant
from the Australian Research Council.
 We are grateful for the computing resources provided
 by the Australian Partnership for Advanced Computing (APAC)
National Facility.



\end{document}